\documentclass[floatfix,amsmath,amssymb,aps,prc,twocolumn,superscriptaddress,showpacs,preprintnumbers,nofootinbib]{revtex4-2}
\usepackage{graphicx,color}
\usepackage{multirow}
\usepackage{times}
\usepackage{bm}
\usepackage{epstopdf}
\usepackage{enumerate}
\usepackage{ulem}
\usepackage{hyperref}
\usepackage{cancel}
\usepackage{slashed}
\usepackage{comment}
\newcommand{\Remove}[1]{{\color[rgb]{0.8,0.3,0.3}\ifmmode\text{\sout{$#1$}}\else\sout{#1}\fi}}
\graphicspath{{./Figs/}}
\makeatletter
\def\@fnsymbol#1{\ensuremath{\ifcase#1\or \dagger\or \ddagger\or
   \mathsection\or \mathparagraph\or \|\or **\or \dagger\dagger
   \or \ddagger\ddagger \else\@ctrerr\fi}}
    \makeatother

\begin{document}

\title{Enhanced Dilepton production near the color superconducting phase and the QCD critical point}
\author{Toru Nishimura}
\affiliation{Department of Physics, Osaka University, Toyonaka, Osaka 560-0043, Japan}
\affiliation{Yukawa Institute for Theoretical Physics, Kyoto University, Kyoto, 606-8502 Japan}

\author{Yasushi Nara}
\affiliation{
Akita International University, Yuwa, Akita-city 010-1292, Japan}
%\affiliation{Frankfurt Institute for Advanced Studies, 
%D-60438 Frankfurt am Main, Germany}
\author{Jan Steinheimer}
\affiliation{Frankfurt Institute for Advanced Studies, Ruth-Moufang-Str. 1, D-60438 Frankfurt am Main, Germany}

\date{\today}
%\preprint{YITP-23-78, KUNS-2968}

\begin{abstract}
The dilepton production yields in relativistic heavy ion collisions are investigated along 
isentropic trajectories in the quark (Wigner) phase within the two-flavor Nambu–Jona-Lasinio model. 
An enhancement of the ultra-low energy dilepton yield in the vicinity of 
the color superconducting (CSC) phase and the QCD critical point (QCD-CP) 
is found, compared to the free quark gas. 
Furthermore, we have found a nontrivial structure in the beam energy dependence 
of the ultra-low energy dilepton yield. 
A local maximum and minimum of the dilepton yield as a function of entropy per baryon 
emerge when the trajectories are close to both locations 
of the CSC phase transition line and the QCD-CP. 
Only the maximum appears in the scenario without the CSC phase but with the QCD-CP. 
On the other hand, when only the CSC phase is considered, 
the dilepton yield monotonically increases as the beam energy decreases. 
These distinctive patterns could potentially serve as signals of the CSC phase and QCD-CP.
In addition, it is found that the dilepton production yield and the location of the minimum strongly 
depend on the value of diquark coupling, suggesting the possibility that the value 
of the diquark coupling may be extracted from experimental data.
\end{abstract}

\maketitle

\section{Introduction}

Probing the QCD phase structure at finite temperature and baryon density, 
realized in high-energy heavy-ion collisions and the core of neutron stars, 
is one of the most critical challenges in nuclear physics and 
astrophysics~\cite{luo2022properties,Fukushima:2010bq,Fukushima:2013rx,Baym:2017whm}.

Heavy-ion collision experiments at top beam energies at the RHIC and the LHC
create high temperature and near-zero baryon density QCD matter, 
which is expected to have existed shortly after the Big Bang~\cite{Heinz:2000bk,Muller:2012zq}. 
The lattice QCD calculations confirm the crossover transition 
from hadronic matter to a quark-gluon plasma state 
at zero baryon density~\cite{Aoki:2006br,Aoki:2006we,Borsanyi:2020fev}.
At high baryon densities, it is expected that there exists the first-order phase transition 
and the QCD critical point (QCD-CP)~\cite{Asakawa:1989bq,Halasz:1998qr} 
as well as the color superconducting (CSC) phase~\cite{Barrois:1977xd,Alford:2007xm} 
from various theoretical studies and signatures for them in heavy-ion collision experiments 
are proposed~\cite{luo2022properties,Lovato:2022vgq,Sorensen:2023zkk}.
To explore the structure of the QCD phase diagram at finite temperatures and baryon densities, 
heavy-ion collision experiments with various collision energies, such as
the beam-energy scan program at RHIC (BES)~\cite{Bzdak:2019pkr}, 
NA61/SHINE~\cite{Snoch:2018nnj}, and HADES~\cite{Galatyuk:2019lcf} are being performed.
Future experiments at FAIR~\cite{Agarwal:2022ydl}, 
J-PARC-HI~\cite{J-PARCHeavy-Ion:2016ikk,Ozawa:2022sam}, and HIAF~\cite{Zhou:2022pxl} are planned. 

As electromagnetic probes decouple from the strongly interacting matter,
it is expected that the lepton pairs produced in the early stages 
of heavy-ion collision carry information about the structure of the QCD matter~\cite{Rapp:2009yu}. 
The effects of chiral mixing in the dilepton emission 
at LHC energy are studied in Ref.~\cite{Sakai:2023fbu}.
The enhancement of dilepton yield from a phase transition 
in heavy ion collision at BES energies is predicted in Ref.~\cite{Savchuk:2022aev}.
In this paper, we consider the dilepton production from virtual photons 
due to the soft mode of the phase transition of the CSC (CSC-PT) and QCD-CP.
In Refs.~\cite{Nishimura:2022mku,Nishimura:2023oqn}, 
it has been investigated how the dilepton production rate is affected by the associated soft modes, 
which are the low-energy collective modes developed 
by the fluctuations of the order parameters around the critical points.
For the CSC-PT, the soft mode arises from the diquark fluctuations 
\cite{Kitazawa:2001ft,Kitazawa:2003cs,Kitazawa:2005vr,Voskresensky:2003wd},
whereas the soft mode of the QCD-CP is formed by a particle-hole collective excitation 
that includes the mixing of baryon number density and energy density 
\cite{Fujii:2003bz,Fujii:2004jt,Yokota:2016tip,Yokota:2017uzu,Son:2004iv}.
It is found that fluctuations of order parameters associated with the CSC-PT and QCD-CP 
can enhance the dilepton production rate in the low-energy/momentum region around 
the respective critical region~\cite{Nishimura:2022mku,Nishimura:2023oqn}.
Creating low-temperature and high-density matter in heavy-ion collisions may be difficult.
However, precursory phenomena of CSC and/or QCD-CP may be observed in heavy-ion collisions
as discussed in Refs.~\cite{Nishimura:2022mku,Nishimura:2023oqn}.

In order to clarify that these dilepton enhancements 
can serve as a signal of the CSC phase or QCD-CP, 
it is necessary to examine the dilepton yields integrated 
over the entire space-time volume along the trajectory in heavy-ion collisions. 
Doing this in a fully dynamic model, which includes a chiral critical point 
in an EoS that incorporates hadronic and quark degrees of freedom, critical dynamics, 
and full non-equilibrium effects, is not yet achieved and is a major challenge in the field. 
In order to justify the efforts that would be required to develop such a description 
we intend to present a qualitative and partially quantitative estimate of 
a possible dilepton signal from the CSC-PT and QCD-CP in heavy ion collisions.
For this purpose, we estimate the excess of the dilepton production yields
over the free quark gas within the two-flavor Nambu–Jona-Lasinio model (NJL)
assuming an isentropic expansion trajectory
in heavy-ion collisions~\cite{Motornenko:2019arp,Motta:2020cbr}.
In this paper, both CSC-PT and QCD-CP are considered 
for the first time in the estimation of dilepton yields.
We shall show that nontrivial structures appear in the beam energy dependence 
of the dilepton yields at ultra-low-energy regions caused by the effects of the CSC-PT and QCD-CP.
We will also examine the sensitivity of the dilepton yield to the diquark coupling.

\section{Model}

To describe the CSC-PT and QCD-CP in a single consistent approach, we use the two-flavor NJL model:
\begin{align}
    \mathcal{L} =\bar{\psi}i(\slashed{\partial}-m_0)\psi  % \nonumber\\
    &+ G_S [(\bar{\psi}\psi)^2  + (\bar{\psi}i\gamma_5\tau\psi)^2] \nonumber\\
    &+ G_D\sum_{A=2,5,7} |\bar{\psi} i \gamma_5 \tau_2 \lambda_A \psi^C|^2
    % (\bar{\psi} i \gamma_5 \tau_2 \lambda_A \psi^C) (\bar{\psi}^C i \gamma_5 \tau_2 \lambda_A \psi),
    % &+ G_D\sum_{A=2,5,7}(\bar{\psi} i \gamma_5 \tau_2 \lambda_A \psi^C)(\bar{\psi}^C i \gamma_5 \tau_2 \lambda_A \psi),
  \label{eq:LC}
\end{align}
where $\psi^C (x) = i \gamma_2 \gamma_0 \bar{\psi}^T (x)$.
The matrices $\tau_2$ and $\lambda_A$ $(A=2,5,7)$ are the antisymmetric components 
of the Pauli and Gell-mann matrices for the flavor SU(2)$_f$ and color SU(3)$_c$, respectively.
The scalar coupling constant $G_S=5.5\, \rm{GeV^{-2}}$, 
the three-momentum cutoff $\Lambda=631\, \rm{MeV}$, and the current quark mass $m_0=5.5\,\mathrm{MeV}$ are
determined to reproduce the pion mass $m_\pi=138\,{\rm MeV}$ 
and the pion decay constant $f_{\pi}=93\, \rm{MeV}$ in vacuum~\cite{Hatsuda:1994pi}.
The magnitude of the diquark coupling $G_D$ has been estimated as $G_D/G_S=0.75$
based on the Fierz transformation of the one gluon exchange interaction in Ref.~\cite{Buballa:2003qv}. 
However, there are large uncertainties in determining the value of the diquark coupling.
We examine the dilepton production with the parameter range 
of $G_D/G_S = 0.6$--$0.7$ considering the following reasons.
For the small values of $G_D / G_S \lesssim 0.5$, the temperature of CSC-PT is very low 
($T_c \approx 10-20\,\text{MeV}$), which may be hardly reached in high-energy heavy-ion collisions. 
On the other hand, for the larger values of $0.75 \lesssim G_D/G_S$, the temperature of CSC-PT 
becomes higher than that of QCD-CP, and the QCD-CP is located inside the CSC phase.
In this case, we have to consider the mixing of the diquark condensate fluctuations 
arising from the CSC-PT and the chiral condensate fluctuations 
from the QCD-CP, which is beyond the scope of this study.
Nevertheless, we will use the low critical temperature parametrization 
to be able to showcase both effects, the CSC and QCD-CP, simultaneously. 
One should, therefore keep in mind that the actual beam energy necessary 
to reach the true QCD-CP (if it exists) may be significantly different from what we will present.

The dilepton production rate at the temperature $T$ and the baryon chemical potential $\mu$ 
per a phase space volume $d^4xd^4k$ is given in terms of the retarded photon self-energy 
$\Pi^{\mu\nu}(\bm{k}, \omega)$ 
for the photon four-momentum $k=(\bm{k}, \omega)$ 
as~\cite{McLerran:1984ay,Weldon:1990iw,Kapusta:1991qp},
\begin {equation}
  \frac{d^8R(k,T,\mu)}{d^4x d^4k}  =  -\frac{\alpha}{12\pi^4}
  \frac{1}{k^2} \frac{1}{e^{\omega/T}-1}
  {\rm Im} {{\Pi}^\mu}_\mu (k,T,\mu) 
       \label{eq:DPR}
\end {equation}
with the fine structure constant $\alpha\approx 1/137$.
In Refs.~\cite{Nishimura:2022mku,Nishimura:2023oqn},
the soft modes near the CSC-PT and QCD-CP 
are incorporated in the retarded photon self-energy, and it has been demonstrated that
the soft modes of the CSC-PT and QCD-CP enhance the production rate 
of dileptons in the low-energy/momentum region.
We utilize the same retarded photon self-energy as in 
Refs.~\cite{Nishimura:2022mku,Nishimura:2023oqn}
(refer to their references for a detailed calculation method of 
the photon retarded self-energy including the respective soft modes).
These photon self-energies are employed to compute the yields of dilepton production 
from three types of contributions to the dilepton production yields:
the soft mode of the CSC-PT and QCD-CP, as well as the free quark gas.
Here, the term `free quark gas' denotes the system of quarks that do not
interact strongly but only electromagnetically.
We note that the effects of the finite energy gap (diquark condensate) 
$\Delta=-2G_D(\bar{\psi}^C i \gamma_5 \tau_2 \lambda_A \psi)$ is not considered 
for our calculation of the retarded photon self-energy in Eq.~\eqref{eq:DPR}.
Incorporating such an effect into our calculations is left as future work.

The (phase) transitions due to spontaneous chiral symmetry breaking and CSC-PT
are calculated in the mean-field approximation.
The pressure in the mean-field approximation is given by~\cite{Kitazawa:2002jop}
\begin{align}
p&=-\frac{M_D^2}{4G_S}-\frac{|\Delta|^2}{4G_D}\nonumber\\
 & +4\int_0^\Lambda \frac{d^3p}{(2\pi)^3}
     \Bigl\{
   E + T \log(1+e^{-\beta\xi_-}) (1+e^{-\beta\xi_+})  \nonumber\\
 & + \epsilon_- + \epsilon_+ 
 + 2T \log(1+e^{-\beta\epsilon_-}) (1+e^{-\beta\epsilon_+})
     \Bigr\},
\end{align}
where $M_D = -2G_S\langle\bar{\psi}\psi\rangle$, $E = \sqrt{p^2+(m_0+M_D)^2}$,
$\xi_\pm = E\pm \mu$,  $\epsilon_\pm = \text{sgn} (\xi_\pm) \sqrt{\xi_\pm^2 + |\Delta|^2}$.
The entropy density and the baryon number density can be obtained from the thermodynamic relations,
$s =  \partial p / \partial T$ and
 $n_B = (\partial p / \partial \mu) / 3$.

\begin{figure}[t]
\centering
\includegraphics[keepaspectratio, scale=0.45]{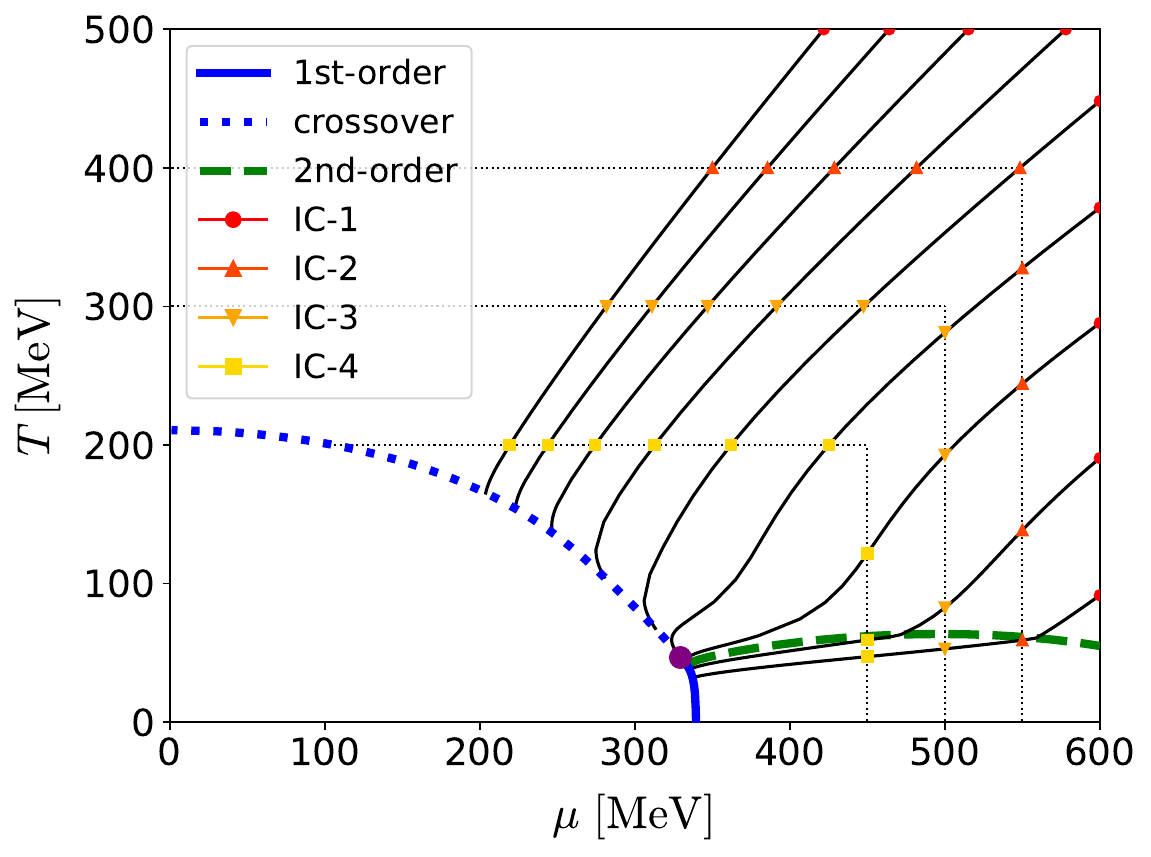}
\caption{
Phase diagram calculated by the two-flavor NJL model~\eqref{eq:LC}
in the mean-field approximation and isentropic lines.
The solid and dotted lines are the first-order and crossover phase transition
of the chiral symmetry breaking, respectively.
The circle marker is the QCD-CP, and the dashed line is 
the second-order phase transition line of the two-flavor CSC 
for the diquark coupling $G_D = 0.7 G_S$.
The thin solid lines are the isentropic lines 
with $S/A = 10, 9, 8, 7, 6, 5, 4, 3, 2$ and 
the small circles, triangles, inverse triangles, and squares represent
the four different initial conditions for each isentropic trajectory.
}
\label{fig:trajectory}
\end{figure}

To specify some trajectories of a system that can be created in heavy-ion collisions,
we assume the Bjorken expansion of the perfect fluid, 
in which the longitudinal component of the fluid velocity is the same as 
the free streaming of particles from the origin: $v_z=z/t$~\cite{Bjorken:1982qr}, 
where $t$ and $z$ are the time and coordinate in the longitudinal direction.
Within the Bojroken ansatz, the time evolution of 
the entropy density and baryon density are given as
\begin {align}
  s (T (\tau), \mu (\tau)) &= s (T (\tau_0), \mu (\tau_0)) \frac{\tau_0}{\tau},
       \label{eq:conservation-s} \\
  n_B (T (\tau), \mu (\tau)) &= n_B (T (\tau_0), \mu (\tau_0)) \frac{\tau_0}{\tau},
       \label{eq:conservation-n}
\end {align}
where $\tau = \sqrt{t^2 - z^2}$ is the proper time 
and $\tau_0$ is the initial proper time.
The expansion is described at lines of constant entropy per baryon, 
$S/A = s(T (\tau), \mu (\tau)) / n_B (T (\tau), \mu (\tau))$.

Figure~\ref{fig:trajectory} shows the isentropic trajectories 
in the $(T,\mu)$ plane together with the chiral-PT line, the QCD critical point, 
and the second-order CSC-PT line obtained from the NJL model.
The isentropic trajectories at different $S/A$ are shown by the thin lines.
To estimate the dilepton yield, we integrate the dilepton production rate along the trajectory 
up to the boundary: either the crossover transition or the first-order phase transition line, 
neglecting the contributions from the hadronic phase but including the CSC phase.
For the smaller $S/A$, the trajectory can go through the CSC phase 
until it reaches the first-order phase transition line. 

We consider four initial conditions (IC), represented by the four types of markers in Fig. 1. IC-1 is defined 
by the horizontal line at $T = 500\, \text{MeV}$ and the vertical line at $\mu = 600\, \text{MeV}$, 
where we denote by $\{T,\mu\}=\{500,600\}$.
In the same way, IC-2, IC-3, and IC-4 are specified by
$\{T,\mu\}=\{400,550\},\{300,500\}$ and $\{200,450\}$, respectively.
As we will show below, dilepton yields at the ultra-low-energy region
integrated along the isentropic trajectory are relatively insensitive 
to the initial conditions in our order estimate.

For a realistic description of the dynamics near the CP,
the production of entropy due to the non-equilibrium fluctuations is important.
This will change the evolution in the phase diagram
as discussed in ~\cite{Herold:2018ptm,Bumnedpan:2022lma}. 
Since we are currently interested in the qualitative relation of dilepton production 
from the CSC-PT and the QCD-CP, more detailed studies will be done in future works.

We assume the 1+1 Bjorken scaling 
to compute the time evolution of the isentropic trajectories. 
When the transverse expansion is included, the system may cool down faster than 
the 1+1 Bjorken longitudinal expansion, which reduces the total dilepton yields. 
However, we expect that transverse expansion may not drastically suppress dilepton yield enhancement. 
A study of the effects of the transverse expansion 
on the dilepton yields is an interesting future work.

%\begin{comment}
\begin{figure}[t]
\centering
\includegraphics[keepaspectratio, scale=0.4]{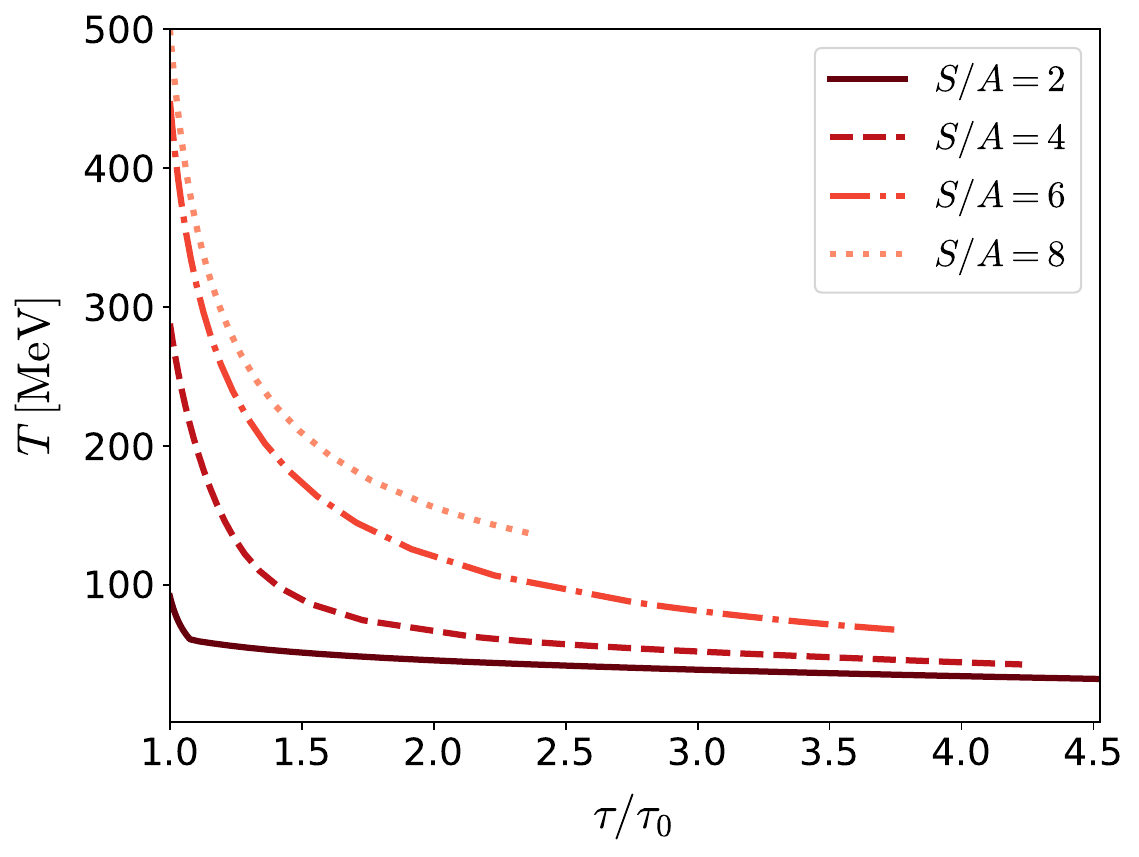}
\includegraphics[keepaspectratio, scale=0.4]{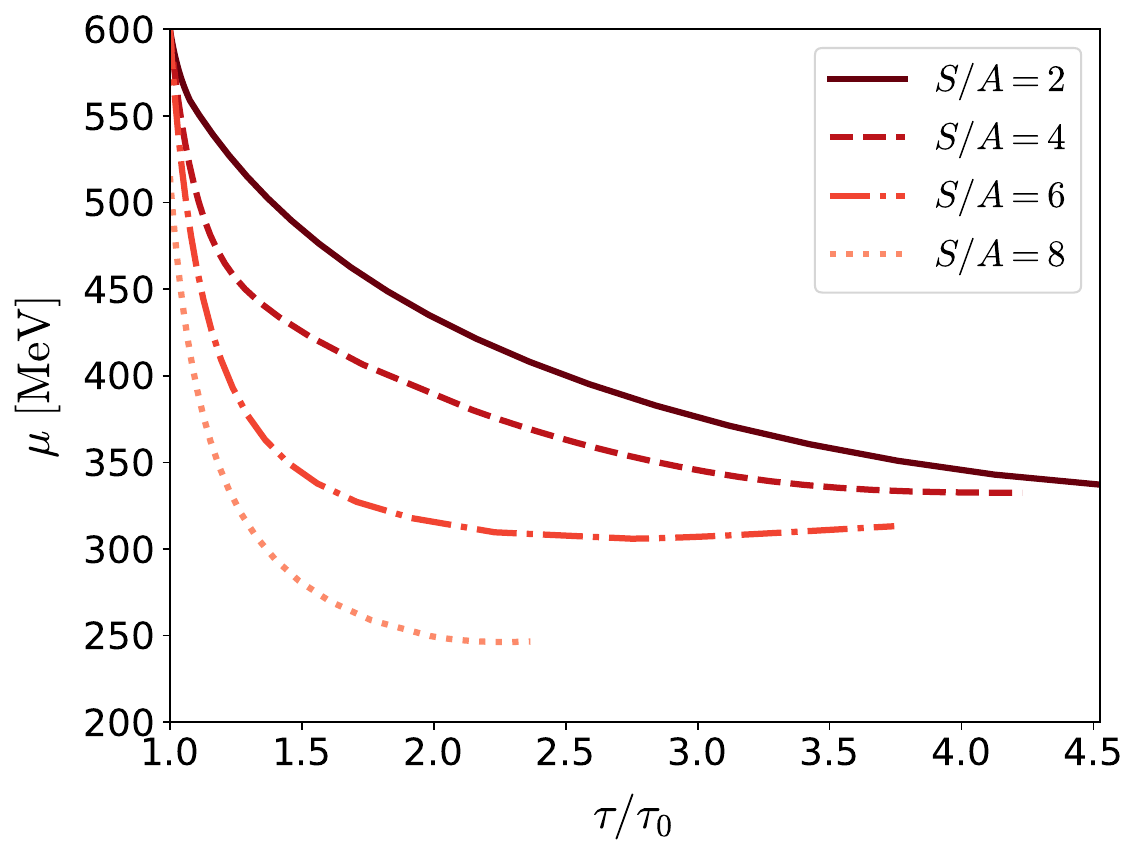}
\caption
{
Time evolution of the temperature $T$ (upper panel) and the baryon chemical potential $\mu$ (bottom panel) 
along the isentropic trajectories for the initial condition 1.
The solid, dashed, dash-dotted, and dotted lines correspond to 
$S/A = 2$, $4$, $6$, and $8$, respectively.
}
\label{fig:tau-dependence_Tmu}
\end{figure}

The time evolution of the temperature and the baryon chemical potential
along the isentropic trajectory as a function of the proper time are 
depicted in Fig.~\ref{fig:tau-dependence_Tmu} for different entropy per baryon number ratios $S/A=2,4,6,8$.
We observe that for larger $S/A$ values, the temperatures and baryon chemical potentials decrease rapidly, 
leading to a shorter lifetime of the system. For smaller $S/A$ values, specifically 
at lower temperatures and higher baryon chemical potentials, 
the time evolution slows down as the system approaches the phase boundary or critical point, resulting 
in a longer lifetime of the system. The lifetime at $S/A=2$ is approximately twice the lifetime at $S/A=8$. 
The extent of enhancement in dilepton production yields depends on 
the proximity of the system's path to the CSC-PT or the QCD-CP.

\section{Results}

In this section, we compute the dilepton yields for different entropy per baryon number $S/A$  
integrated over the isentropic trajectories.
The behavior of the dilepton yield in the energy spectrum is qualitatively similar to the result 
of the invariant spectrum, as confirmed in Refs.~\cite{Nishimura:2022mku,Nishimura:2023oqn}.
Therefore, We consider the energy spectrum with zero momentum for the dilepton production rate.
We also examine the cut-off dependence of the energy of dilepton.

\begin{figure*}[thb]
\begin{tabular}{ccc}
      \begin{minipage}[t]{0.32\hsize}
        \centering
        \includegraphics[keepaspectratio, scale=0.3]{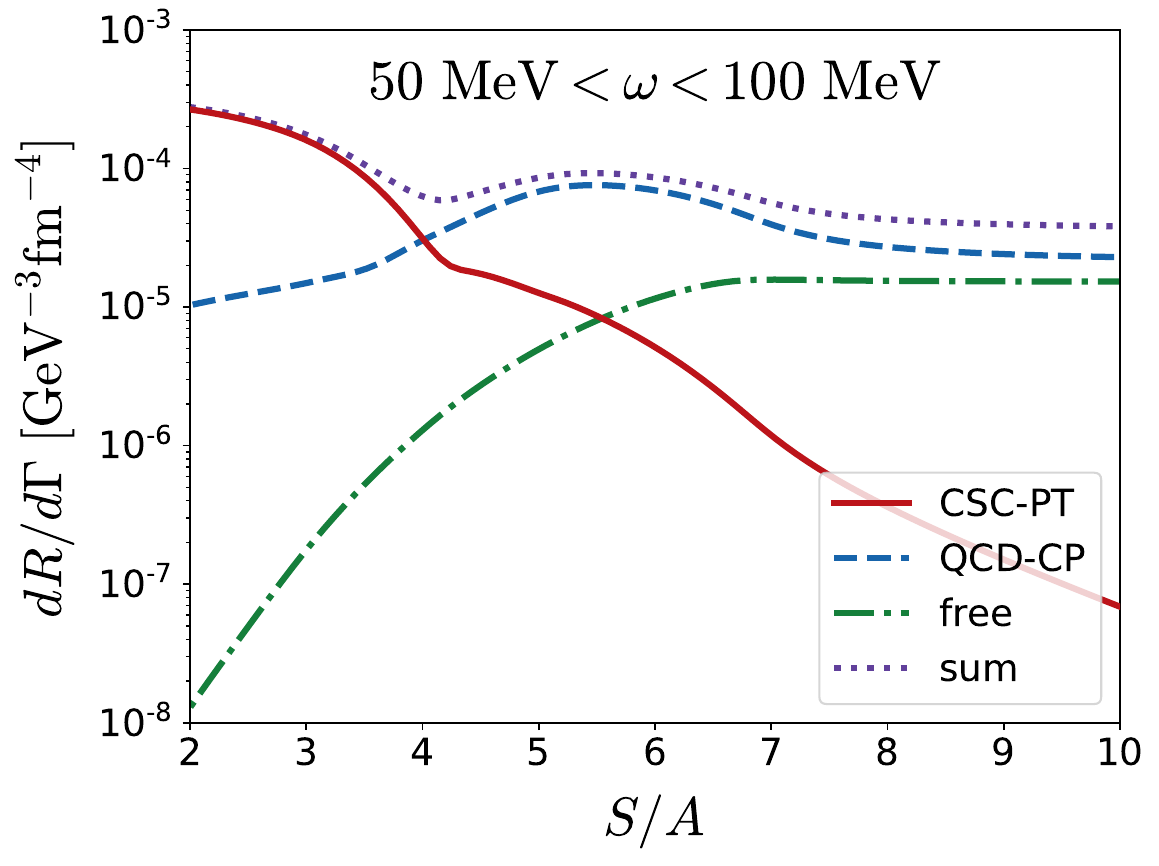}
      \end{minipage} &
      \begin{minipage}[t]{0.32\hsize}
        \centering
        \includegraphics[keepaspectratio, scale=0.3]{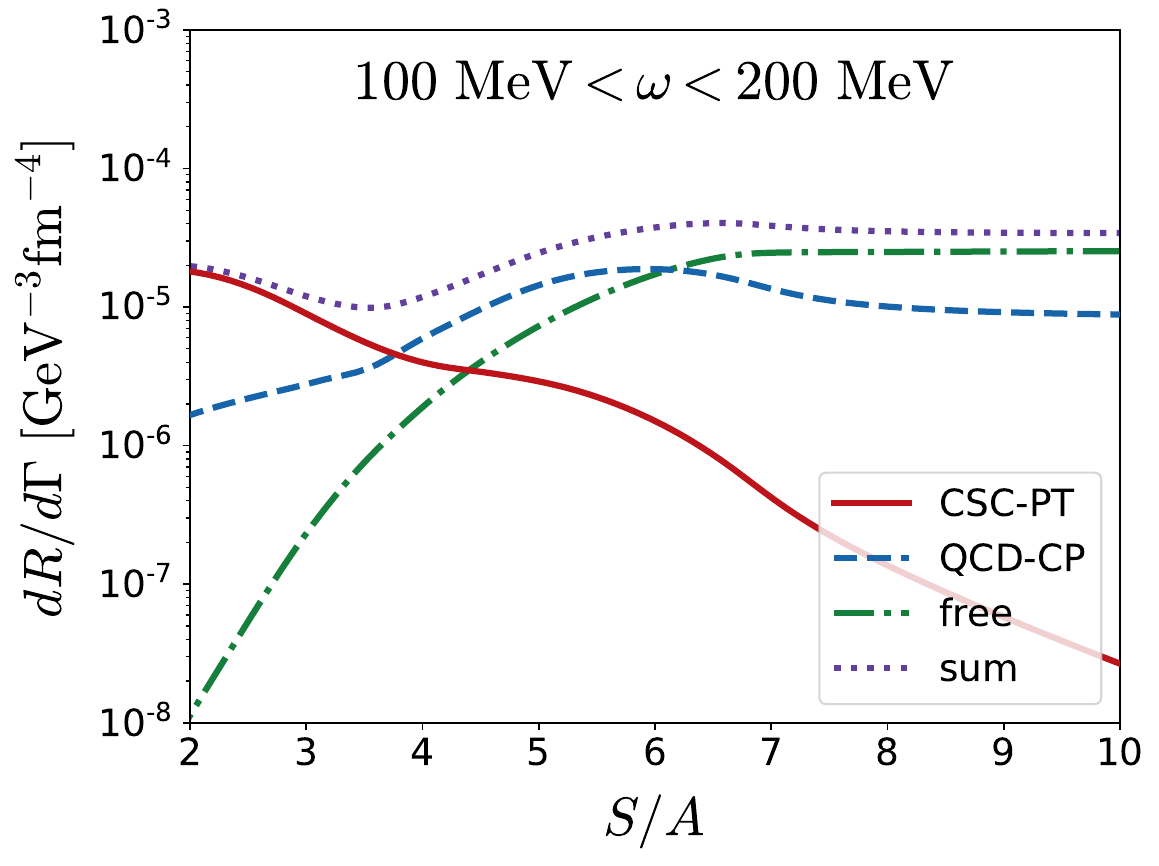}
      \end{minipage} &
      \begin{minipage}[t]{0.32\hsize}
        \centering
        \includegraphics[keepaspectratio, scale=0.3]{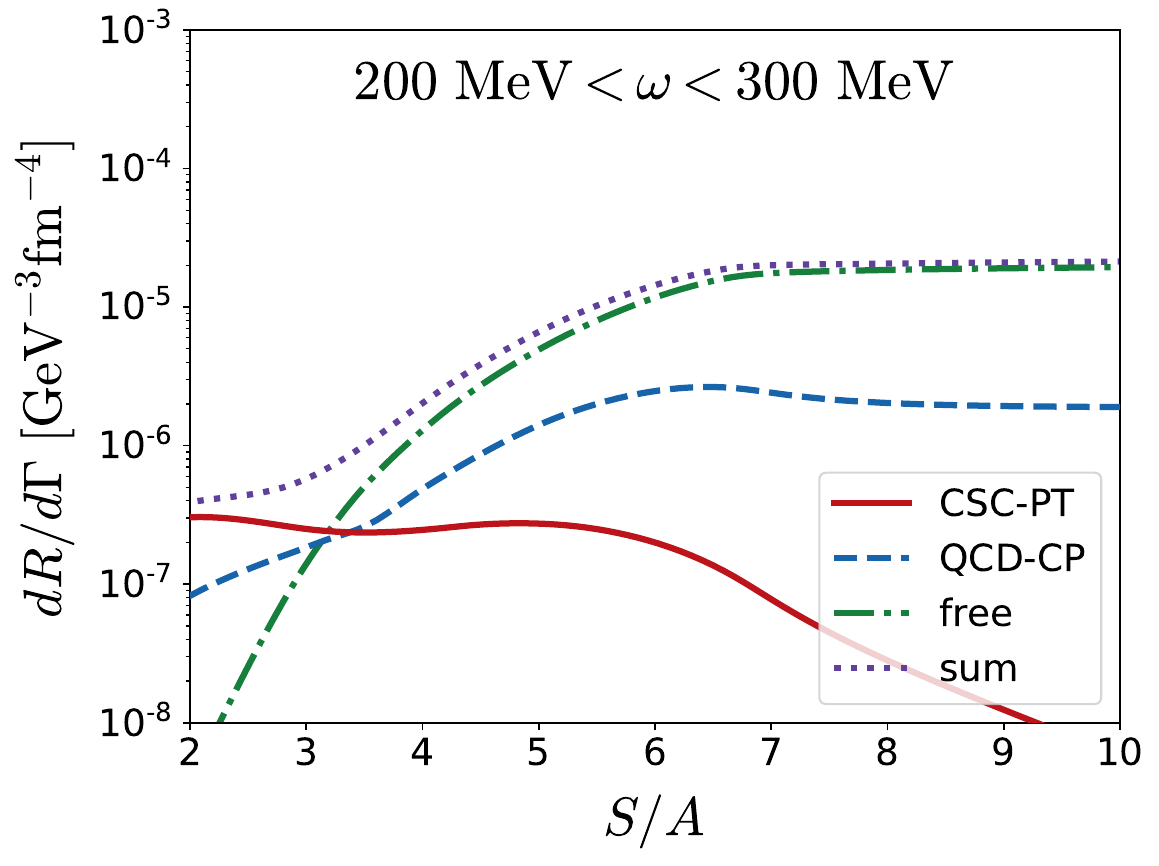}
      \end{minipage} 
\end{tabular}

\begin{tabular}{ccc}
      \begin{minipage}[t]{0.32\hsize}
        \centering
        \includegraphics[keepaspectratio, scale=0.3]{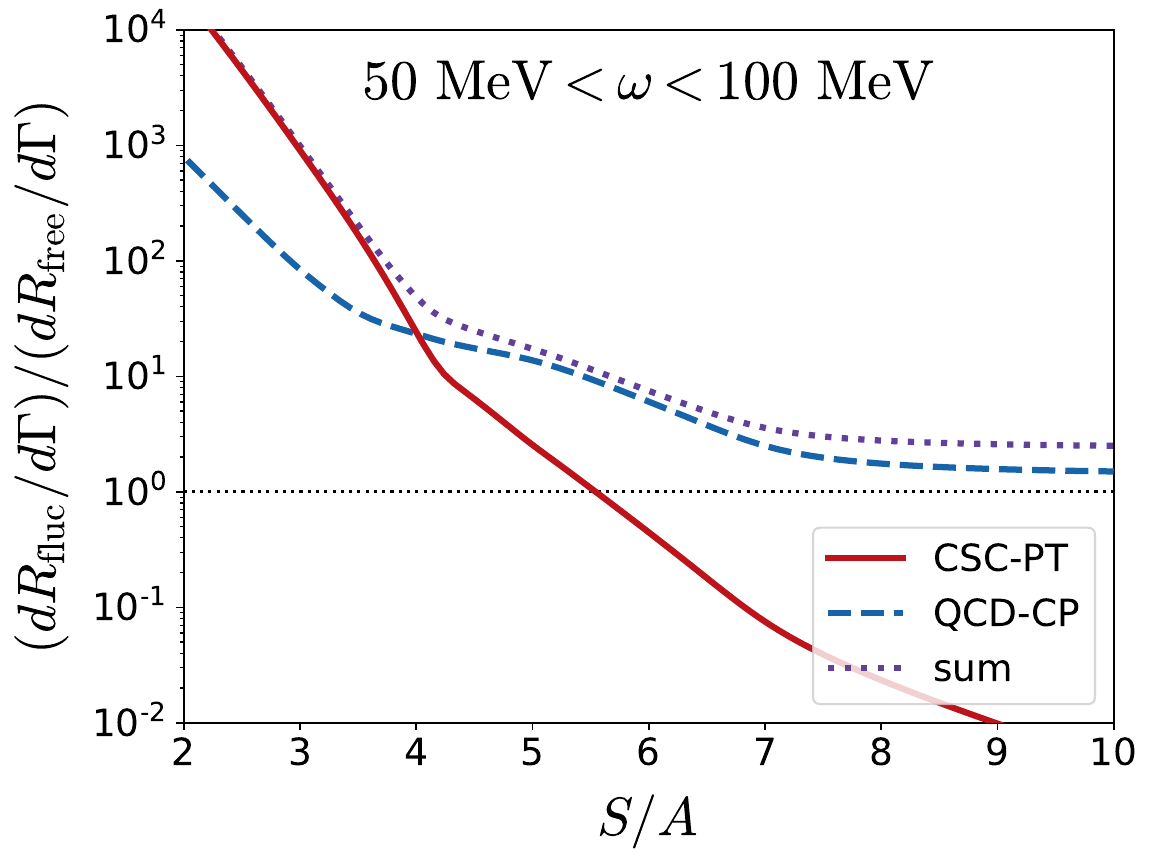}
      \end{minipage} &
      \begin{minipage}[t]{0.32\hsize}
        \centering
        \includegraphics[keepaspectratio, scale=0.3]{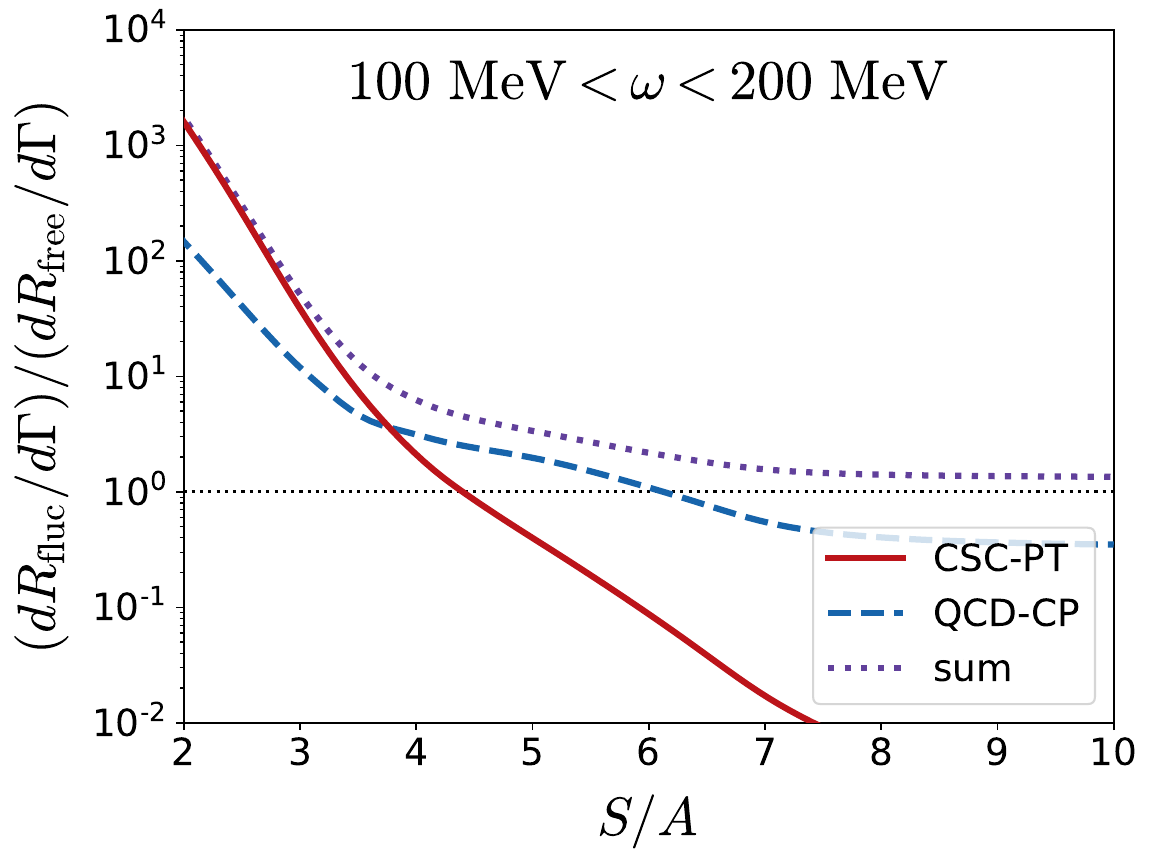}
      \end{minipage} &
      \begin{minipage}[t]{0.32\hsize}
        \centering
        \includegraphics[keepaspectratio, scale=0.3]{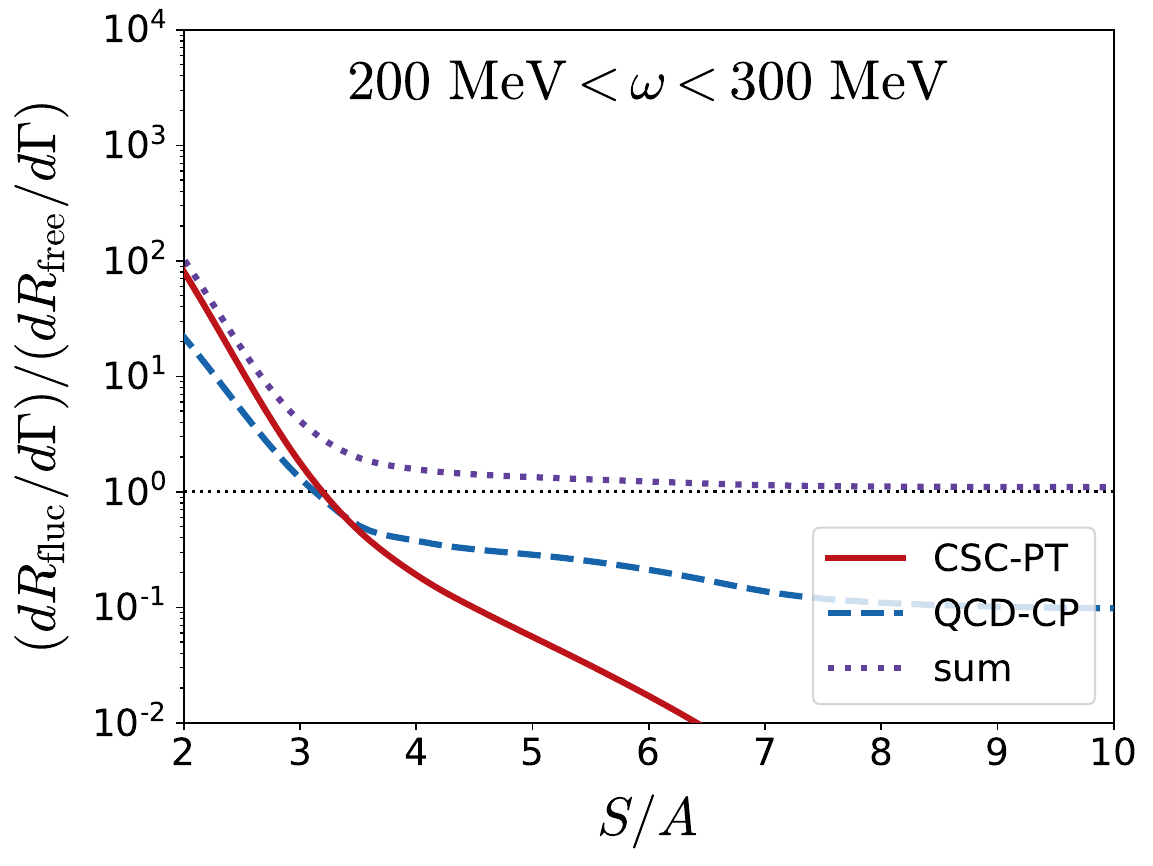}
      \end{minipage} 
\end{tabular}

\begin{tabular}{ccc}
      \begin{minipage}[t]{0.32\hsize}
        \centering
        \includegraphics[keepaspectratio, scale=0.3]{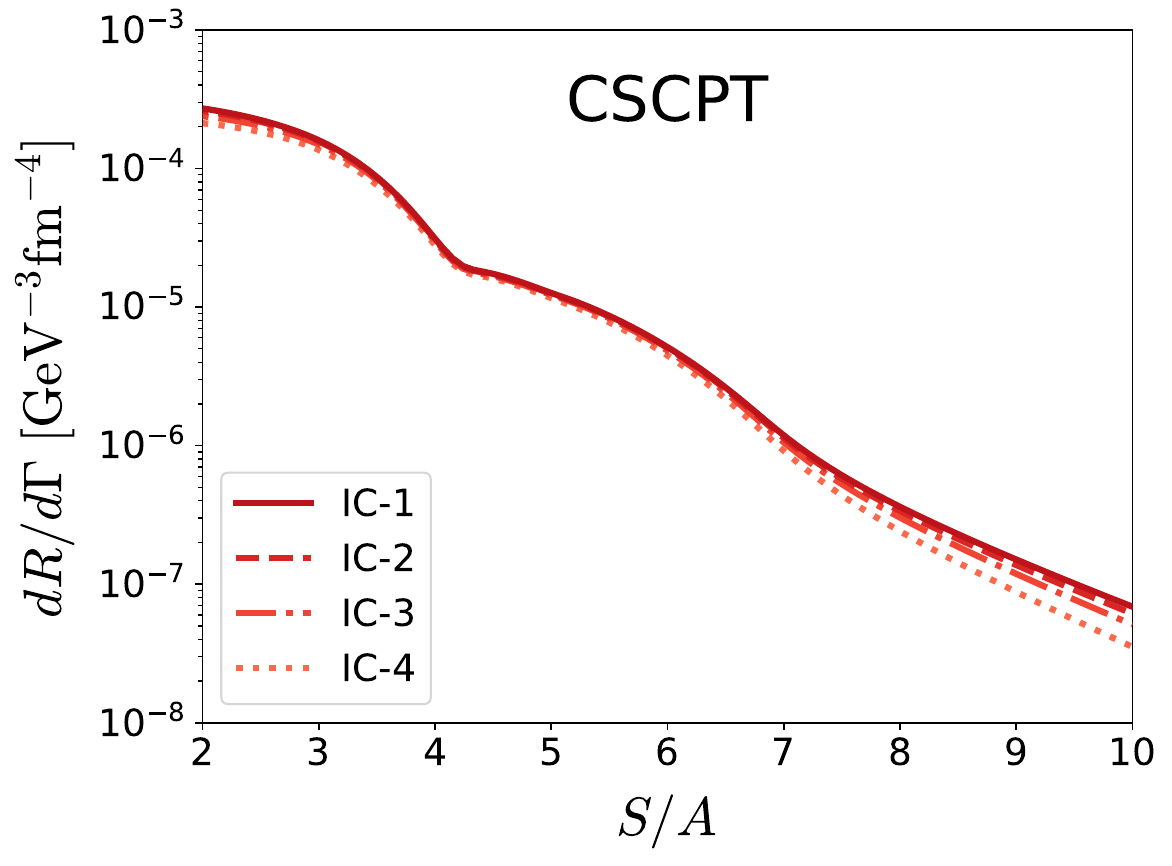}
      \end{minipage} &
      \begin{minipage}[t]{0.32\hsize}
        \centering
        \includegraphics[keepaspectratio, scale=0.3]{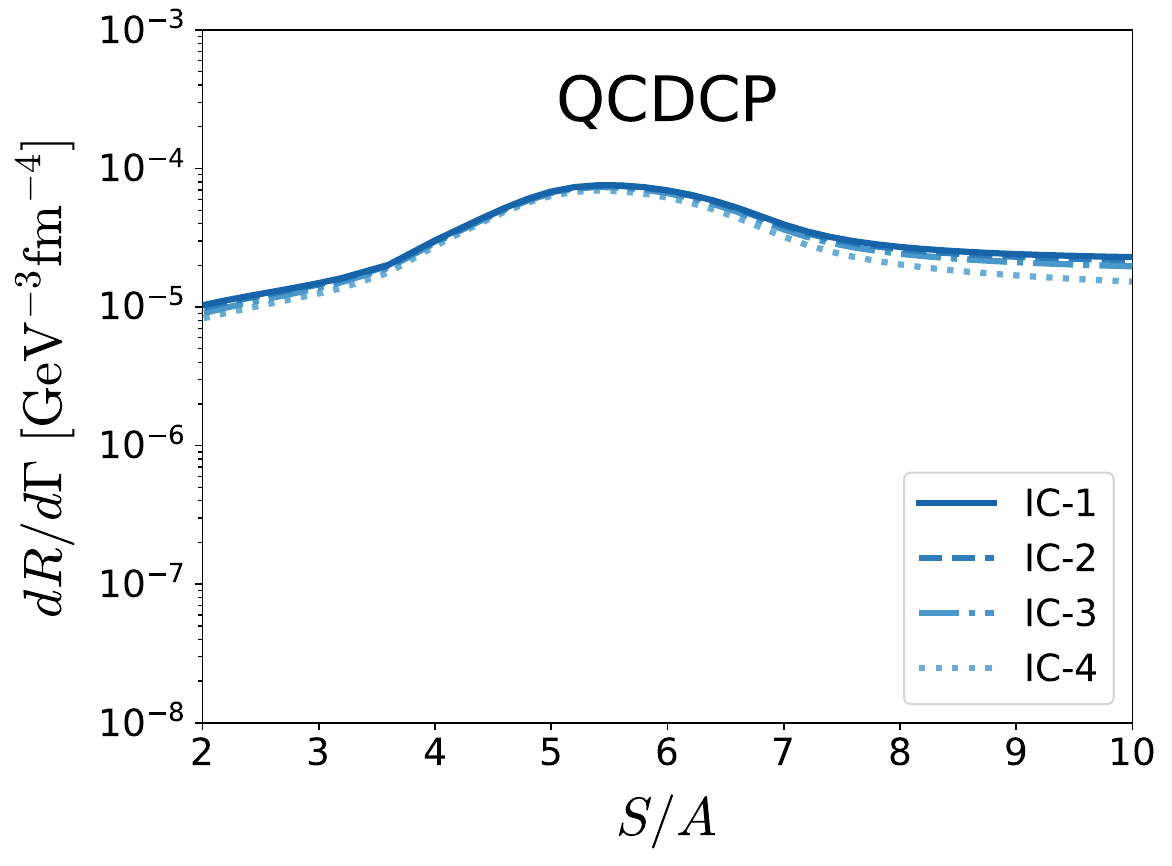}
      \end{minipage} &
      \begin{minipage}[t]{0.32\hsize}
        \centering
        \includegraphics[keepaspectratio, scale=0.3]{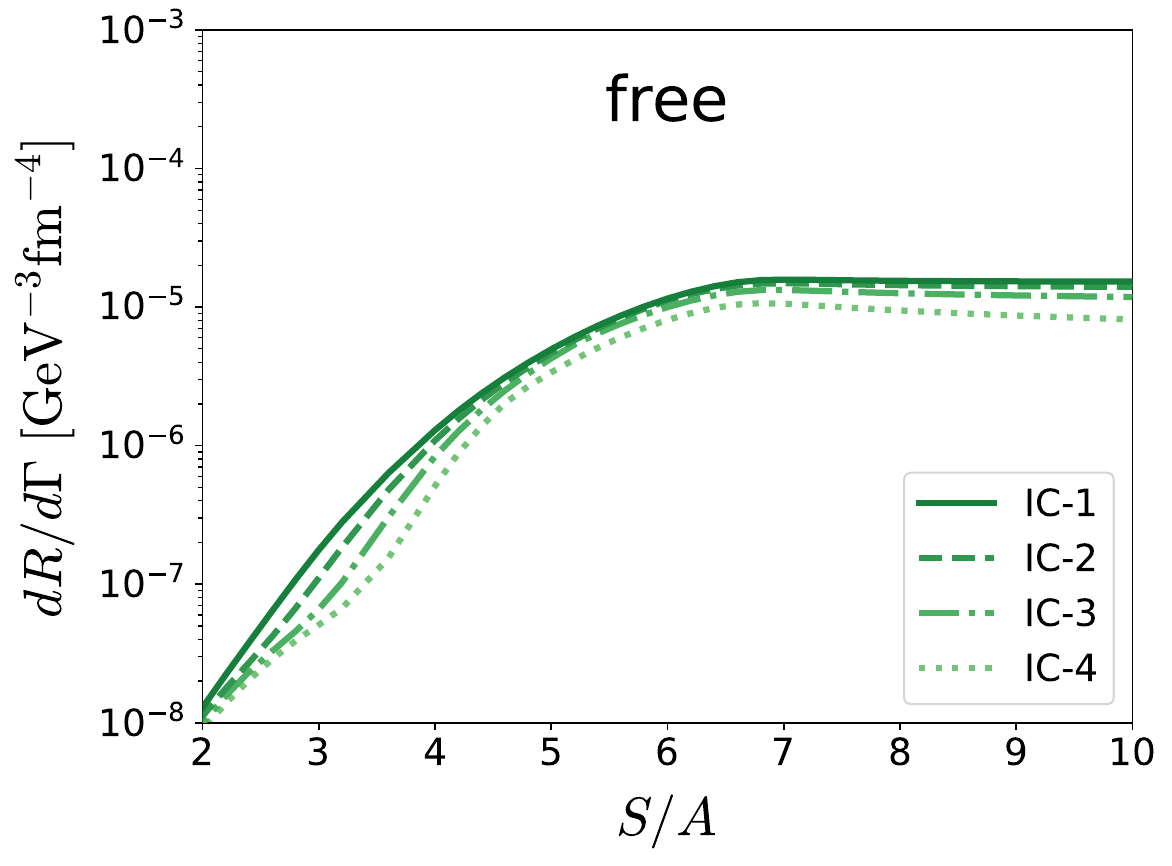}
      \end{minipage} 
\end{tabular}
\caption{
Dilepton yields integrated over an isentropic trajectory for the initial condition 1
as a function of the entropy to baryon number ($S/A$) (top panels)
are shown for different integral regions of the dilepton energy: The left panel shows 
the results for $50\,\text{MeV}<\omega <100\,\text{MeV}$,
while mid panel is for $100\,\text{MeV}<\omega <200\,\text{MeV}$,
and right panel is for $200\,\text{MeV}<\omega <300\,\text{MeV}$.
In the mid panels, the ratio of the dilepton yields (mid panels)
are plotted. The integration ranges of the mid-three panels are the same
as those for the top panels.
The initial condition dependence of the dilepton yields 
for $50\,\text{MeV}<\omega <100\,\text{MeV}$are shown in the bottom panels.
The value of diquark coupling $G_D = 0.7 G_S$ is used for all calculations.
}
\label{fig:RATE}
\end{figure*}

\begin{figure*}[thb]
\begin{tabular}{ccc}
      \begin{minipage}[t]{0.32\hsize}
        \centering
        \includegraphics[keepaspectratio, scale=0.31]{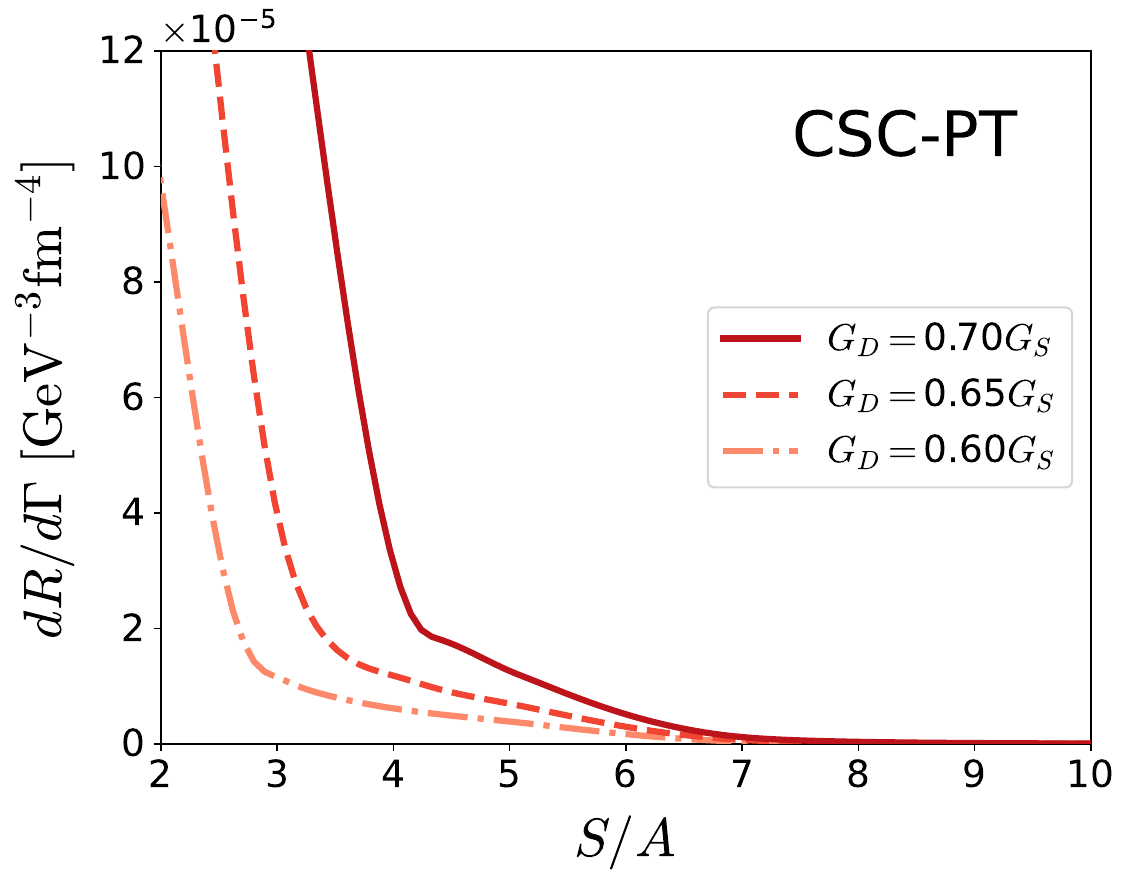}
      \end{minipage} &
      \begin{minipage}[t]{0.32\hsize}
        \centering
         \includegraphics[keepaspectratio, scale=0.31]{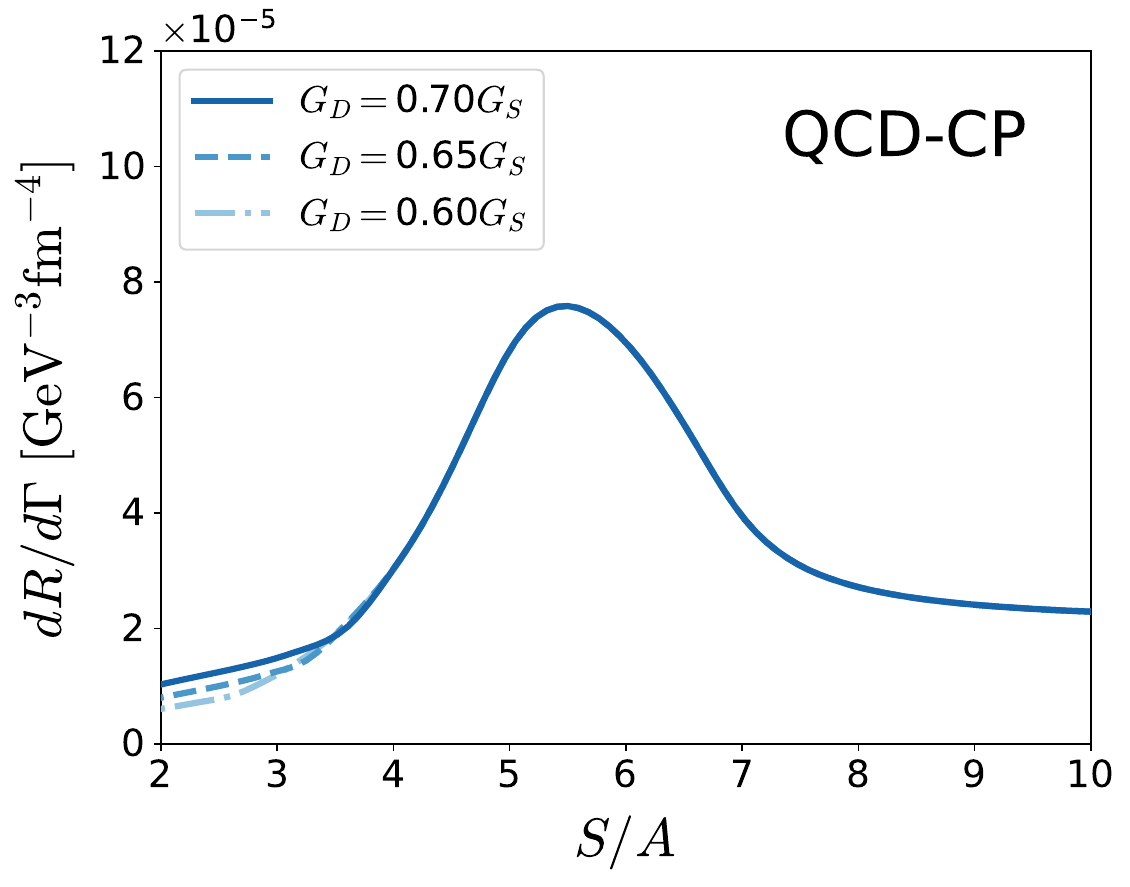}
      \end{minipage} &
      \begin{minipage}[t]{0.32\hsize}
        \centering
        \includegraphics[keepaspectratio, scale=0.31]{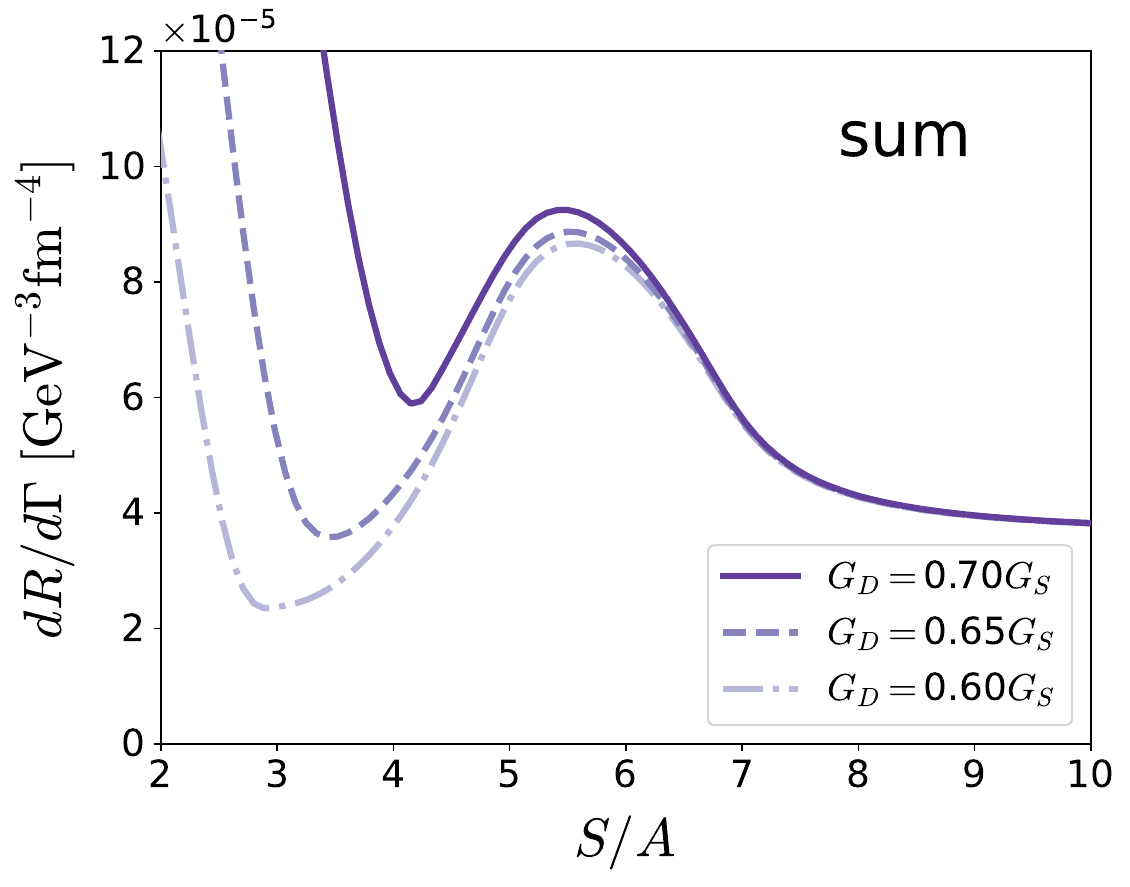}
      \end{minipage} 
\end{tabular}
\caption{Entropy per baryon ratio dependence of the dilepton yields integrated 
over an isentropic trajectory for the initial condition 1
are shown for different values of the diquark coupling $G_D$.
The integration range of the dilepton energy is taken as $50\,\text{MeV}<\omega <100\,\text{MeV}$.}
\label{fig:RATE_GD}
\end{figure*}

In the top panels of Fig.~\ref{fig:RATE}, we show 
the dilepton production at $\bm{k} = \bm{0}$ as a function of $S/A$
\begin {align}
\frac{dR}{d\Gamma}\equiv
\frac{1}{\tau_0^2}\frac{d^6R}{dx dy d\eta d^3k}
= \int_{C (S/A)} \frac{\tau d\tau}{\tau_0^2} \int d\omega 
\frac{d^8R(\bm{0}, \omega)}{d^4x d\omega d^3k},
\end {align}
where the integral path $C$ is taken along the constant $S/A$
and the value of diquark coupling $G_D = 0.7 G_S$ is used. 
In the top and middle panels of the figure, 
the IC-1 is used as an initial condition.
Instead of introducing a model for the initial conditions,
the dilepton yields are divided by the factor $\tau_0^2$ as we are only concerning 
the enhancement of the dilepton production from the fluctuations
of soft mode relative to the contributions from the free quark gas.
Three ranges of integration are considered for $\omega$-integral
(left: $50 < \omega/{\rm MeV} < 100$, 
middle: $100 < \omega/{\rm MeV} < 200$, 
and right: $200 < \omega/{\rm MeV} < 300$).
The solid red lines represent the contributions of CSC-PT, 
the dashed blue lines represent the contributions of QCD-CP, 
and the dotted green lines correspond to the yields from the free quark gases.
The sum of them gives the total dilepton production. 
We find that the dilepton production yield from the soft modes 
in the low-energy region ($50 < \omega/{\rm MeV} < 100$)
exceeds the yield of the free quark gases in a wide range of $S/A$ values.
In the dilepton energy range of $100 < \omega/{\rm MeV} < 200$,
we observe the enhancement of the dilepton at $S/A < 4$ for both QCD-CP and CSC.

A local maximum and minimum in the dilepton yields are observed 
at $S/A \approx 5.5$ and at $S/A \approx 4.0$, respectively.
These bumps appear because the system's trajectory approaches close to the QCD-CP 
at $S/A \approx 5.5$ or largely aligns with the CSC-PT
at $S/A \approx 4.0$, as illustrated in Fig.~\ref{fig:trajectory}. 

In the second row panels of Fig.~\ref{fig:RATE},
we plot the ratio of the dilepton production from the CSC-PT and QCD-CP 
to the free quark gas. Dilepton enhancement over the free quark gas 
is seen in the region of $\omega / \,\text{MeV} < 300$ and $S/A < 3-4$. 
The enhancement for very low energy is very significant, several orders of magnitude.
$S/A<4$ corresponds to the collision energy of $\sqrt{s_{NN}} \leq 3.0\,\text{GeV}$
according to the estimate in Ref.~\cite{Motornenko:2019arp}. 
However, due to our simplifications and the model-dependence of the CP location, 
we do not want to claim that we expect quark matter or the CSC phase to appear at such a low beam energy. 

We now check the dependence of the dilepton yields on different initial conditions, 
i.e., the expansion time in the quark phase.
The initial conditions for low- to intermediate energies may be calculated, e.g.
by the Rankine-Hugoniot-Taub adiabat~\cite{Motornenko:2019arp}.
Instead of using specific models for initial conditions, we shall show that
enhancement of dilepton yields at ultra-low energy regions is not strongly
affected by the details of the initial conditions.
In the bottom panels of Fig.\ref{fig:RATE}, we show the dilepton yields 
for $50 < \omega / \,\text{MeV} <100$ with four different initial conditions 
for each trajectory where four different initial conditions 
are specified as `IC-1,2,3,4' in Fig.~\ref{fig:trajectory}.
It is seen that dilepton production yields from each contribution, 
CSC-PT, QCD-PT, and free quark gas are not affected largely 
by the initial conditions in a wide range of the $S/A$ values.
We anticipate that the maxima will persist even when incorporating the contributions from the CSC phase. 
This is simply due to the incredibly large effect of the CSC and QCD-CP on the observable yield in our approach. 
A small change in the free quark contribution due to an extended lifetime 
is not relevant when compared to a several-order-of-magnitude enhancement.

As mentioned earlier, the $T$- and $\mu$-dependence in the invariant mass spectrum 
is qualitatively similar to the energy/momentum spectrum
as shown in Refs.~\cite{Nishimura:2022mku, Nishimura:2023oqn}
Thus, we expect that the structure in the $S/A$ dependence for the dilepton production 
from the fluctuations of the CSC-PT and QCD-CP can still be seen in the invariant mass spectra.

As the boundary of the CSC phase is sensitive to the diquark coupling,
it would be interesting to examine the diquark coupling dependence.
The diquark coupling dependence for the dilepton yields is shown in Fig.~\ref{fig:RATE_GD}.
The effect of QCD-CP appears as the local maximum at $S/A\approx 5.5$.
The local minimum is due to the soft modes of CSC-PT.
The dilepton yield and the location of the local minimum are sensitive to the diquark coupling, 
which suggests that heavy-ion collision experiments may provide information on the diquark coupling.

We observe that the sharp increase in the dilepton yield from the CSC-PT contribution 
at lower $S/A$ stems from the neglect of the finite energy gap effects 
in our calculation of the retarded photon self-energy. 
This abrupt rise is prominent when the trajectory is within the CSC phase. 
We anticipate that incorporating the effects 
of the finite energy gap will suppress this sharp increase. 
We further expect that the results for QCD-CP and the free quark gas
would not change much when the effects of the finite energy gap are considered 
in the retarded photon self-energy~\cite{Nishimura:2024}.

\section{Summary and discussion}

We have computed the dilepton yields in ultra-low energy regions within the two-favor NJL model 
assuming a simplified longitudinal expansion and isentropic trajectories
to examine the entropy per baryon dependence as a proxy of the beam energy dependence.
It is demonstrated that the dilepton yields are greatly enhanced near 
the color-superconducting phase and the QCD critical point in comparison to the free quark gas. 
Furthermore, we predict a local maximum and minimum in the entropy per baryon ratio dependence 
(and thus the beam energy dependence) of ultra-low energy dilepton yields.
We expect that this structure persists for the invariant mass spectra  because 
the behavior of the invariant mass spectra is qualitatively similar to the energy spectra 
of the dilepton yield as shown in Refs.~\cite{Nishimura:2022mku,Nishimura:2023oqn}.
This structure could be a signal of the superconducting phase if the system created 
in heavy-ion collisions crosses the CSC phase boundary or reaches close to QCD-CP,
which may be examined in the current RHIC-BES FIXT target experiment and the HADES as well as 
future experiments FAIR, J-PARC-HI, and HIAF. 
We also found that the dilepton production yield and the location of the minimum due to CSC
is sensitive to the diquark coupling, which suggests that experimental data 
may provide constraints on the value of the diquark coupling.

The magnitude of the enhancement observed is several orders of magnitude. 
One should note, however, that the actual enhancement that will be observed is likely 
reduced due to non-equilibrium effects and finite equilibration times near the QCD-CP. 
The purpose of our study is to establish the effect qualitatively 
and point out certain systematic observables
when the CSC phase and QCD-CP are implemented simultaneously. 
The next step will be to put these effects into a more realistic dynamical description 
of heavy ion collisions at low beam energies with an equation of state 
that also includes a description of the hadronic and nuclear matter phases of QCD.
For instance, the hadronic phase may be computed by connecting the parity doublet model 
with the NJL model~\cite{Minamikawa:2020jfj,Minamikawa:2021fln}.
Non-equilibrium chiral Bjoriken expansion has been 
investigated within the linear sigma model and predicts significant 
entropy production~\cite{Herold:2018ptm,Bumnedpan:2022lma}. Such non-equilibrium effects
should be investigated. In addition as a future work, we plan 
to compute the dilepton yields employing a full 3+1-dimensional fluid dynamical 
simulation~\cite{Huovinen:2002im,Endres:2014zua,Galatyuk:2015pkq,Endres:2015fna,Akamatsu:2018olk}.
Putting all these sophisticated developments in a single model description 
will have a quantitative impact on the enhancement, 
but it is out of the scope of the current paper.
Nevertheless, our results encourage more realistic calculations 
of the dilepton yields in heavy-ion collisions.

\begin{acknowledgments}
T. N. and Y. N. thank the team at the Frankfurt Institute of Advanced
Studies where a part of this work was done for their splendid hospitality during their visit.
T. N. thanks M. Kitazawa, T. Kunihiro, and K. Murase for their valuable comments.
This work was supported in part by the Grants-in-Aid for Scientific Research from JSPS (Nos. JP21K03577), 
JST SPRING (Grant No.~JPMJSP2138) and Multidisciplinary PhD Program for Pioneering Quantum Beam Application. 
J.S. thanks the Samson AG for their support. The authors acknowledge 
for the support of the European Union's Horizon 2020 research 
and innovation program under grant agreement No 824093 (STRONG-2020).
\end{acknowledgments}

%\appendix

\bibliography{ref}

\end{document}